\documentclass[twocolumn, superscriptaddress,longbibliography]{revtex4-1}
\usepackage{blindtext}
\usepackage{algorithm2e}
\usepackage{float}
\usepackage[utf8]{inputenc}
\usepackage[T1]{fontenc}
\usepackage{amsmath}
\usepackage{subfigure}
\usepackage{ulem}
\usepackage{bm,color,bbm}
\usepackage{hyperref, mathtools}
\usepackage{amsfonts}    
\usepackage{graphicx}   
\usepackage{verbatim}   
\usepackage{booktabs}
\usepackage{threeparttable}
\usepackage{dcolumn}
\usepackage{multirow}
\usepackage{array}
\usepackage{upgreek}
\usepackage{pifont}
\usepackage{textcomp}

\begin{document}
\title{Enhancing Deep Learning Based Structured Illumination Microscopy Reconstruction with Light Field Awareness}

\author{Long-Kun Shan}
\affiliation{CAS Key Laboratory of Quantum Information, University of Science and Technology of China, Hefei, 230026, China}
\affiliation{CAS Center For Excellence in Quantum Information and Quantum Physics, University of Science and Technology of China, Hefei, 230026, China}

\author{Ze-Hao Wang}
\email{zehao@ustc.edu.cn}
\affiliation{CAS Key Laboratory of Quantum Information, University of Science and Technology of China, Hefei, 230026, China}
\affiliation{CAS Center For Excellence in Quantum Information and Quantum Physics, University of Science and Technology of China, Hefei, 230026, China}

\author{Tong-Tian Weng}
\affiliation{CAS Key Laboratory of Quantum Information, University of Science and Technology of China, Hefei, 230026, China}
\affiliation{CAS Center For Excellence in Quantum Information and Quantum Physics, University of Science and Technology of China, Hefei, 230026, China}

\author{Xiang-Dong Chen}
\affiliation{CAS Key Laboratory of Quantum Information, University of Science and Technology of China, Hefei, 230026, China}
\affiliation{CAS Center For Excellence in Quantum Information and Quantum Physics, University of Science and Technology of China, Hefei, 230026, China}
\affiliation{Hefei National Laboratory, University of Science and Technology of China, Hefei 230088, China}

\author{Fang-Wen Sun}
\affiliation{CAS Key Laboratory of Quantum Information, University of Science and Technology of China, Hefei, 230026, China}
\affiliation{CAS Center For Excellence in Quantum Information and Quantum Physics, University of Science and Technology of China, Hefei, 230026, China}
\affiliation{Hefei National Laboratory, University of Science and Technology of China, Hefei 230088, China}

\date{\today}

\begin{abstract}
    Structured illumination microscopy (SIM) is a pivotal technique for dynamic subcellular imaging in live cells. Conventional SIM reconstruction algorithms depend on accurately estimating the illumination pattern and can introduce artefacts when this estimation is imprecise. Although recent deep learning–based SIM reconstruction methods have improved speed, accuracy, and robustness, they often struggle with out-of-distribution data. To address this limitation, we propose an Awareness-of-Light-field SIM (AL-SIM) reconstruction approach that directly estimates the actual light field to correct for errors arising from data distribution shifts. Through comprehensive experiments on both simulated filament structures and live BSC1 cells, our method demonstrates a 7\% reduction in the normalized root mean square error (NRMSE) and substantially lowers reconstruction artefacts. By minimizing these artefacts and improving overall accuracy, AL-SIM broadens the applicability of SIM for complex biological systems.

\end{abstract}


\flushbottom
\maketitle
\thispagestyle{empty}
\hfill

\section{Introduction}
    \begin{figure}[ht]
        \centering
        \includegraphics[width=\linewidth]{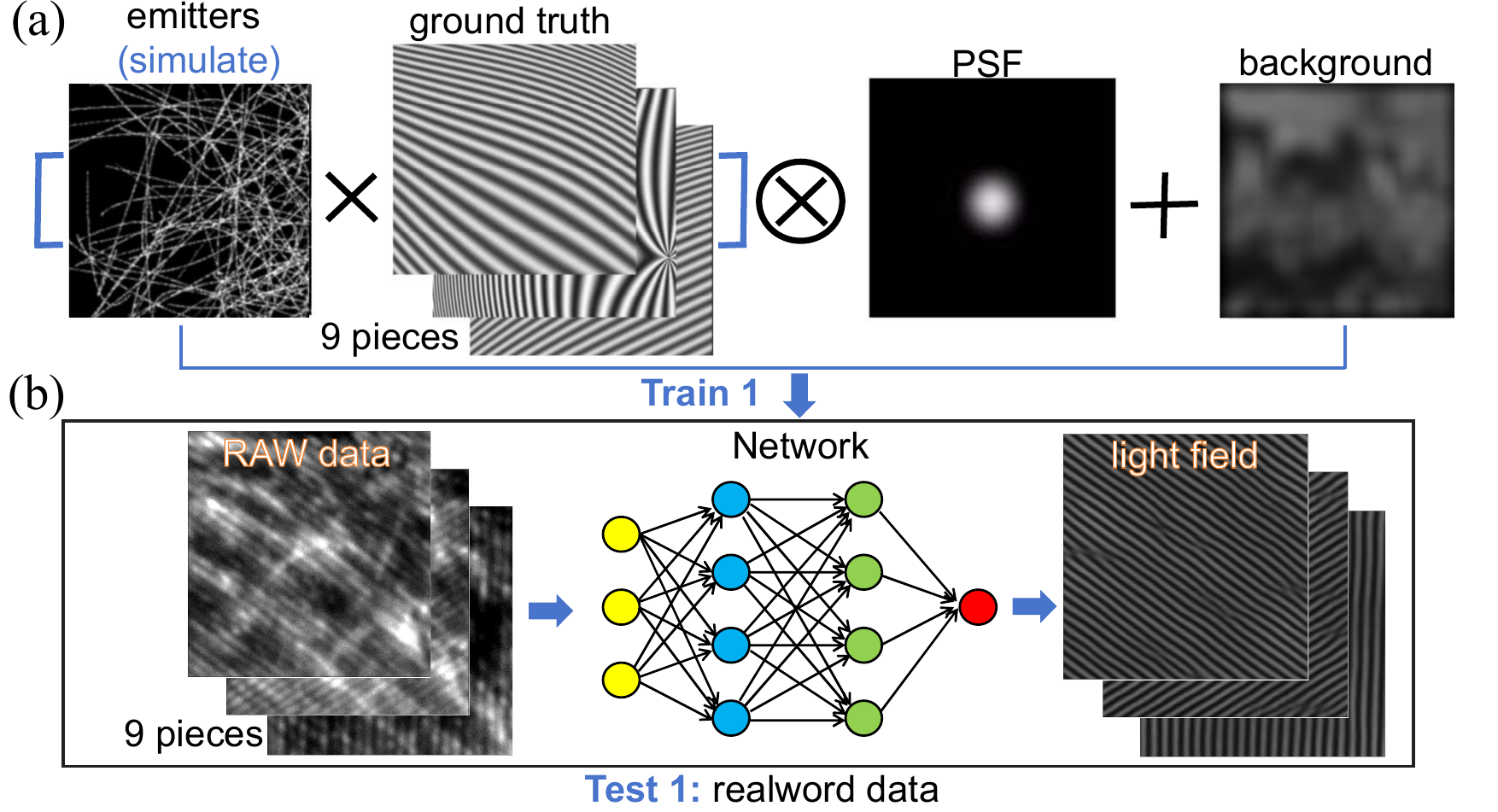}
        \includegraphics[width=\linewidth]{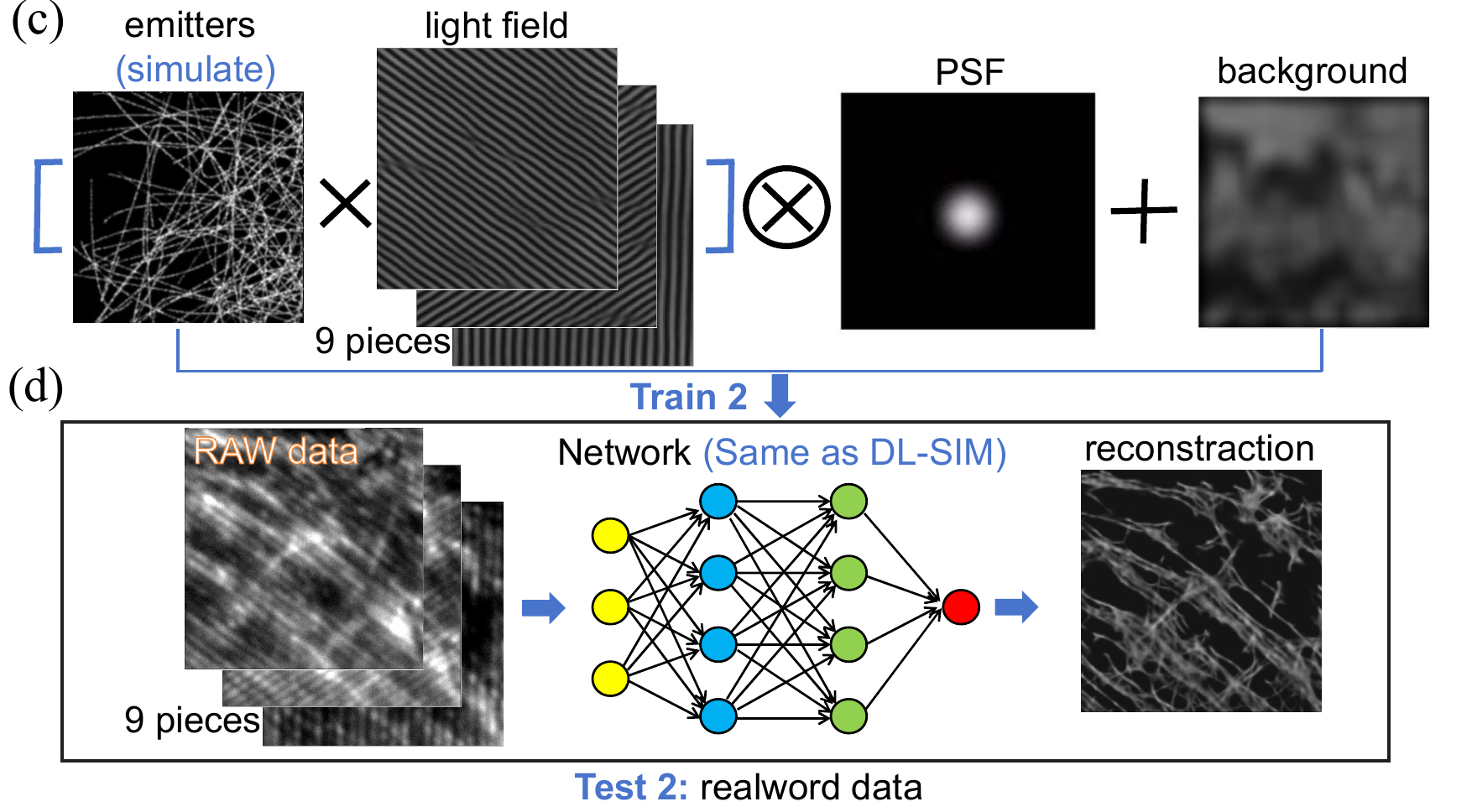}
        \caption{AL-SIM pipeline. (a) The first training stage. (b) the first test stage. (c) the second training stage. (d) the second test stage}
        \label{fig:al_sim}
    \end{figure}
    The Abbe diffraction limit~\cite{AbbeE1873Contributions} confines the spatial resolution of fluorescence microscopy to roughly half the wavelength of light, posing a significant challenge in visualizing subcellular structures in detail. To overcome this constraint, super-resolution microscopy techniques have been developed over the past decades. Noteworthy examples include stochastic optical reconstruction microscopy~\cite{Rust2006Sub, Bates2007Multicolor}, photoactivated localization microscopy~\cite{Hess2006Ultra, Shroff2008Live}, and stimulated emission depletion microscopy~\cite{Klar2000Fluorescence, Betzig2006Imaging}. Among these innovations, structured illumination microscopy (SIM~\cite{Gustafsson2000Doubling}) stands out for its rapid imaging capabilities, broad field of view, compatibility with a variety of fluorescent probes, and the potential to achieve resolutions twice that allowed by diffraction.

    Reconstruction algorithms are essential for structured illumination microscopy (SIM)~\cite{chen2023superresolution} as they extract high-frequency details from raw images captured under structured light illumination. Conventional SIM reconstruction typically employs Fourier transform-based methods that require accurate knowledge of the light field’s physical parameters. However, practical imperfections such as distortion, scattering, and noise can introduce subtle discrepancies, resulting in pronounced artefacts~\cite{mo2021structured, Belthangady2019Applications}.

    Deep learning frameworks have become a powerful tool in SIM, providing super-resolved image reconstruction, enhanced resolution, and reduced artefacts stemming from inaccurate parameter estimation~\cite{chen2023superresolution,christensen2021ml, qiao2021evaluation, qiao2023rationalized}. Recent advancements in DL-SIM highlight its versatility and performance improvements across multiple dimensions. Universal reconstruction models based on transfer learning have been developed, enabling cross-system compatibility and achieving real-time imaging speeds of 200ms per frame\cite{christensen2021ml}. The incorporation of prior knowledge of illumination patterns into rationalized deep learning (rDL) for SIM has enabled sustained live-cell imaging with minimized phototoxicity, significantly enhancing the ability to observe rapid subcellular dynamics\cite{qiao2023rationalized}. Additionally, neural networks have been shown to reduce the required number of raw images by five-fold compared to traditional SIM, while maintaining exceptional photon efficiency for low-light conditions (100× fewer photons)\cite{jin2020deep}. However, DL-SIM encounters a significant hurdle known as the out-of-distribution problem~\cite{shen2021towards}. When light fields or sample structures significantly deviate from the training set, DL-SIM models may produce artefacts or fail to maintain consistent reconstruction quality. This issue arises when a model trained under specific conditions produces artefacts in different scenarios, thereby impeding robust generalization in microscopic imaging~\cite{Belthangady2019Applications}.
    
    In this work, we propose a novel approach called AL-SIM to mitigate the out-of-distribution problem in DL-SIM reconstructions. As illustrated in Fig.~\ref{fig:al_sim}, AL-SIM uses a two-stage algorithm. First, a deep learning model predicts the perturbed illumination light field from the raw SIM data. Next, rather than relying on the standard cosine illumination, we use the predicted light field to generate simulated data for model training, thereby reducing data bias. We validate AL-SIM through comparative experiments on simulated light field distortions and real-world BSC1 cell imaging, demonstrating that it achieves resolution comparable to conventional DL-SIM while significantly reducing artefacts and improving the NRMSE.

\section{Method}
    The AL-SIM method employs a two-stage approach. In the first stage, the model is trained to accurately estimate the light fields from the raw SIM data, aiming to capture a distribution closer to the actual physical light field. In the second stage, the predicted light fields from the first stage are used to generate simulated data that more closely reflects the real distribution. This data is then employed to train a reconstruction model, which is subsequently used to produce the super-resolution image from the raw SIM data. 

    \begin{figure}[ht]
        \centering
        \includegraphics[width=\linewidth]{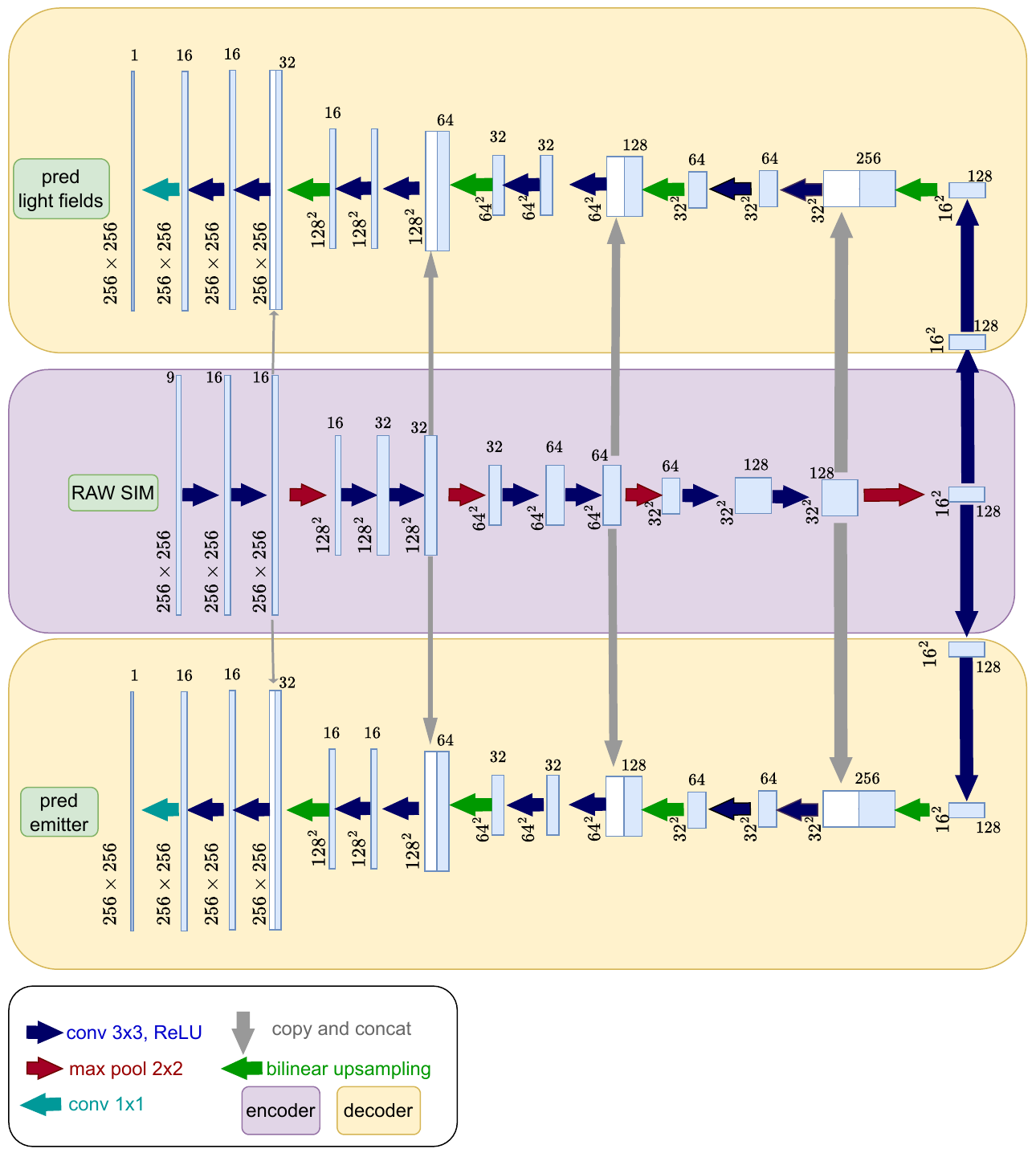}
        \caption{AL-SIM network architecture. The network is based on U-net.}
        \label{fig:unet}
    \end{figure}

    \textbf{AL-SIM pipeline.} Fig.~\ref{fig:al_sim} provides an overview of the AL-SIM pipeline. Figs.~\ref{fig:al_sim}(a–b) illustrate the first training stage, in which a network is optimized to predict light fields. Specifically, simulated cosine light fields with distortions are multiplied by emitters, convolved with the simulated point spread function (PSF), and then combined with simulated noise and background to form a synthetic SIM dataset. As shown in Fig.~\ref{fig:unet}, the network adopts a U-net architecture~\cite{ronneberger2015u}; its encoder processes the raw SIM data, and the two decoders subsequently reconstruct the features to predict both the light fields and the emitters. The loss function is computed by comparing the predicted outputs with the ground truth, thereby guiding the network to accurately estimate light fields from the raw SIM data. These predicted light fields are then used to generate bias-corrected simulated data for the second training stage. Figs.~\ref{fig:al_sim}(c–d) illustrate the second stage, where the light fields predicted from real SIM data are employed to create simulated SIM data. Here, the model is trained using a loss function based on the discrepancy between the predicted emitters and the ground truth. Finally, the trained model is applied to real SIM data, producing bias-corrected super-resolution images.
    
    \textbf{Data Synthesis Methodology.} Our simulated training dataset comprises emitters, light fields, point spread functions (PSF), noise, and background. For emitters, scatter points, curves, and natural-image–based patterns have proven effective and are already applied in DL-SIM~\cite{ling2020fast, zhang2022deep}. Following this approach, we generate random scatter points and curves while using COCO data~\cite{lin2014microsoft} for natural images. The random curve generation algorithm proceeds as follows: (1) randomly add control points within the field of view, with each curve containing 100–200 points for diversity; (2) interpolate a smooth curve through these points using Catmull-Rom splines~\cite{catmull1974class}, parameterized from 0 to 1; (3) apply random brightness variations along the curve, starting at 100 and incremented by values in the range of -2 to 2, to simulate natural fluctuations.

    For the light fields, the simulated cosine pattern adheres to the parameters of our two-beam interference SIM, with the period varying between 327 nm and 546 nm. Additionally, nine simulated light fields incorporate distortions and aberrations to mirror the imperfections observed in real SIM data. First, we define a set of nonlinear functions (such as sine, cosine, and tangent). Then, for each channel of each light field image, we randomly select three of these functions and generate three coefficients that sum to 1. We combine these functions and coefficients into a composite function, which is subsequently applied to distort the nine light field images. This approach yields light field images with substantial diversity and complexity.

    Gaussian noise with a relative scale range of 0.1–1 is introduced to simulate imaging noise. The PSF closely matches parameters from our experimental setup—an objective lens with a numerical aperture of 1.30, a fluorescence wavelength of 600 nm, and a pixel size of 40 nm. A Gaussian filter with $\sigma=16~\text{pixel}$ is used to emulate smooth background variations commonly seen in microscopic imaging. Finally, emitters, light fields, background, the PSF, and noise are integrated to produce the simulated raw SIM data.

    \textbf{Loss Function.} To train AL-SIM, we introduce a composite loss function comprising two terms: one measuring the discrepancy between the predicted light fields and their ground truth, and the other measuring the discrepancy between the predicted emitters and their ground truth. Both terms are computed using the same function \(\hat{L}\), defined as:

    $$
    \hat{L} = \alpha \cdot L_{\text{ms-ssim}} + (1-\alpha) \cdot L_{\text{L1}},
    $$
    
    leading to the overall loss function:
    
    $$
    \text{Loss} = \beta \cdot \hat{L}_\text{emitters} + (1-\beta) \cdot \hat{L}_\text{light fields},
    $$
    
    where \(L_{\text{ms-ssim}}\) and \(L_{\text{L1}}\) represent the multi-scale structural similarity index~\cite{zhao2016loss} and the sum of absolute differences (L1), respectively. In our setup, \(\alpha = \tfrac{1}{8}\) and \(\beta = \tfrac{1}{2}\).
    
    Using both \(L_{\text{ms-ssim}}\) and \(L_{\text{L1}}\) for training offers multiple benefits. The \(L_{\text{ms-ssim}}\) term captures perceptual and structural information aligned with human vision, while the \(L_{\text{L1}}\) term enforces pixel-wise accuracy by minimizing the sum of absolute differences. This combination has proven particularly effective for image restoration~\cite{zhao2016loss}.

    \begin{figure}[ht]
        \centering
        \includegraphics[width=\linewidth]{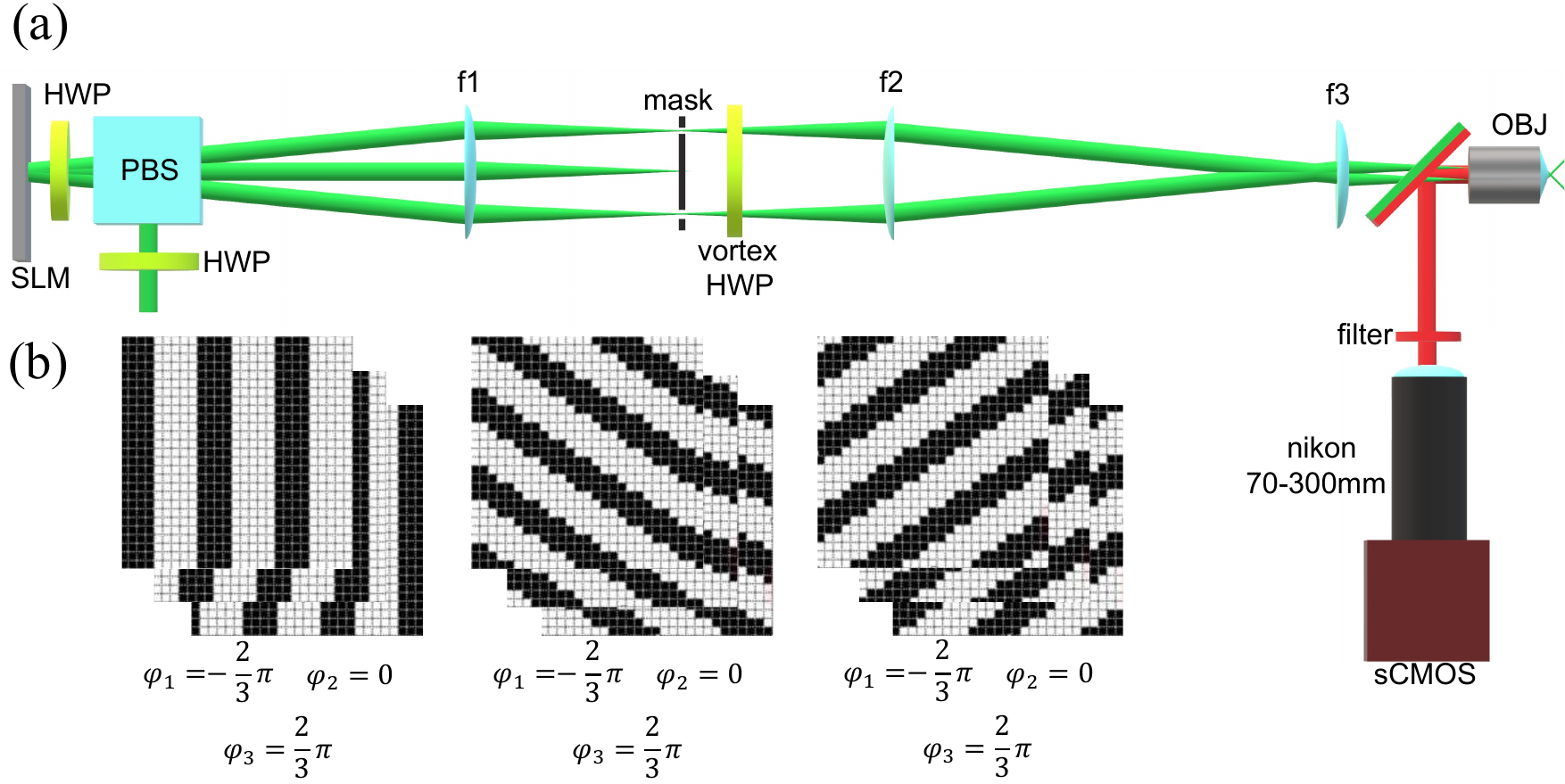}
        \caption{Experimental setup. (a) Optical path of SIM (b) Formation of different illumination light fields and phase modulation }
        \label{fig:optical_path}
    \end{figure}

    \textbf{Conventional SIM Algorithm.} The conventional SIM algorithm operates by separating high- and low-frequency components in the raw data, which is collected under different phases of the illuminating light field. In the Fourier domain, high-frequency information is shifted to the zero-frequency position and combined with low-frequency components, thereby expanding the overall frequency spectrum. Sinusoidal illumination in multiple directions further broadens the spectrum along different orientations, enhancing spatial resolution. In our experiment, we use fairSIM~\cite{muller2016open} as the traditional SIM reconstruction algorithm.

    \textbf{DL-SIM.} In our implementation of DL-SIM, we employ the widely adopted U-Net architecture for supervised super-resolution tasks to ensure generalizability.  The network comprises an encoder-decoder structure with skip connections, where both the encoder and decoder consist of four hierarchical levels. At each level, the encoder applies two convolutional layers followed by downsampling, while the decoder performs upsampling followed by two convolutional layers. The number of filters in the convolutional layers increases progressively from 32 to 64, 128, and finally 256, enabling the network to extract and refine features at different spatial scales. The model takes nine-channel SIM raw images as input and outputs single-channel emitter predictions. For fairness in comparison, the second-stage emitter prediction model in AL-SIM utilizes the identical U-Net architecture as DL-SIM.

    \textbf{Double-Beam Interference Super-Resolution Structured Light Microscope.} As illustrated in Fig.~\ref{fig:optical_path}(a), a 532\,nm laser is expanded by a 5X beam extender to uniformly illuminate the active region of a ferroelectric liquid crystal on silicon spatial light modulator (FLCOS-SLM; 2048\(\times\)1536 pixels, QXGA-3DM, Forth Dimension Displays). A polarizing beam splitter (PBS) and a half-wave plate (HWP) are placed between the laser and the FLCOS-SLM to produce linearly polarized light.  

    As shown in Fig.~\ref{fig:optical_path}(b), a binary grating pattern is loaded onto the FLCOS-SLM. The FLCOS-SLM, PBS, and HWP together form a polarization grating that concentrates the laser energy onto the \(\pm\)1 diffraction orders. Other orders are blocked by a mask, and the zero-order vortex half-wave plate (vortex HWP) rotates the polarization direction of the \(\pm\)1 beams to generate S-polarization, thereby maximizing the contrast of the structured illumination. After traveling through a 4\(f\) system, the \(\pm\)1 diffracted beams interfere on the sample surface to form structured light.  

    Fluorescence emitted by the sample is split off via a dichroic mirror and recorded by an sCMOS camera (Hamamatsu, C13440-20CU) through a long-pass filter. Nine grating patterns with varying directions and phases (Fig.~\ref{fig:optical_path}(b)) generate distinct structured light fields. The super-resolution fluorescence image is then reconstructed from nine raw images acquired under these different illumination patterns.
    
    \textbf{Cell Culture.} African green monkey kidney epithelial cells (BSC1) are cultured in DMEM supplemented with 10\% fetal bovine serum (FBS). The cells are maintained under standard conditions (5\% \(\text{CO}_2\), humidified atmosphere at 37\textcelsius) and seeded onto No.~1.5 glass-bottom dishes 24~hours before sample preparation. For cell passage, the cells are washed three times with pre-warmed PBS and then digested with 25\% trypsin.

    Cells are grown on 35\,mm No.~1.5 glass coverslips in glass-bottom dishes. On the day of sample preparation, they reach 50\%–70\% confluence and are fixed for 10~minutes in pre-warmed (37\textcelsius) fixation buffer containing 4\% EM-grade paraformaldehyde and 0.1\% glutaraldehyde in PBS. Following fixation, the samples are washed three times with 2\,mL PBS and incubated for 30~minutes at room temperature in PBS containing 5\% BSA and 0.5\% Triton~X-100. Actin stain~555 (Cytoskeleton, PHDH1-A), diluted in the same BSA–Triton solution, is applied, and the cells are incubated for more than 12~hours at 4\textcelsius in the dark. After incubation, the samples are washed three times with 2\,mL PBS for 5~minutes each. A post-fixation buffer is then applied for 10~minutes, followed by five washes in sterile water. The remaining sterile water is removed, and the samples are air-dried, sealed with parafilm, and stored at -20~\textcelsius.

\section{Result}
\begin{figure}[ht]
        \centering
        \includegraphics[width=\linewidth, 
                    trim=0 0 0 3cm, 
                    clip]{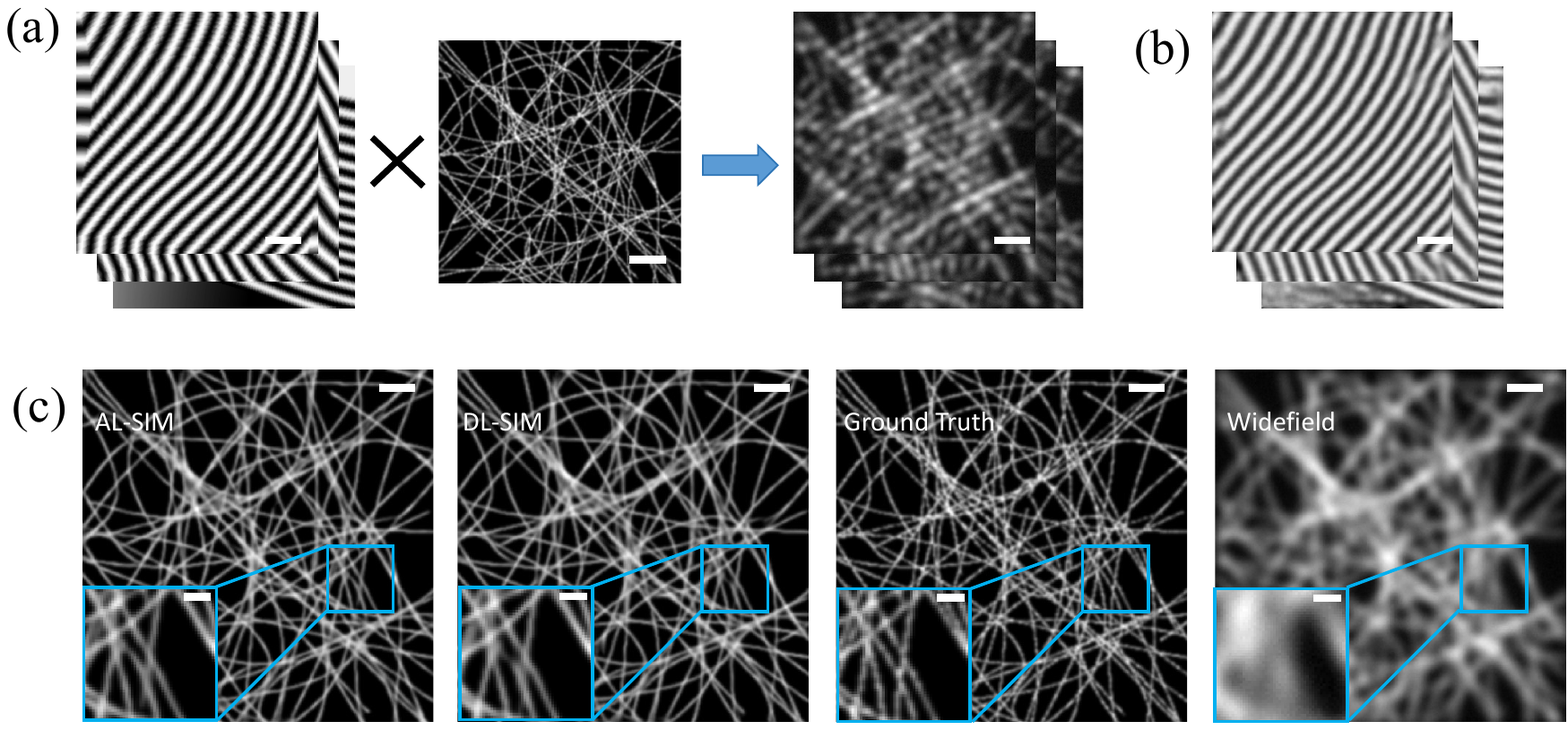}
        \caption{Simulation validation of AL-SIM under light field distortions. 
        (a) Distorted illumination patterns (left), simulated filament structures (middle), and raw SIM images (right). 
        (b) Illumination patterns predicted by the AL-SIM first-stage model.
        (c) Comparative reconstruction results: (i) AL-SIM, (ii) DL-SIM, (iii) Ground Truth, (iv) Widefield microscopy. 
        Scale bars: 1\,\textmu m (main panels), 200\,nm (magnified insets)}
        \label{fig:result-simulate}
    \end{figure}

    We perform simulation experiments to verify AL-SIM by simulating SIM imaging of filament structures under light field distortions. To test its generalisation, we introduce a distortion distribution different from the one used during training. Fig.\ref{fig:result-simulate}(a) presents a representative example from the test set, illustrating the distorted light field, the underlying filament structures, and the raw images. We generate 100 test images and train and evaluate both DL-SIM and AL-SIM. Fig.\ref{fig:result-simulate}(b) displays one of the inferred light fields, which will subsequently be utilized for stage 2 training of AL-SIM, to develop an emitter prediction model. Finally, DL-SIM achieves an average NRMSE of 0.068, while AL-SIM achieves 0.063. As shown in Fig.~\ref{fig:result-simulate}(c), AL-SIM significantly reduces artefacts compared to DL-SIM.

\begin{figure}[ht]
        \centering
        \includegraphics[width=0.9\linewidth]{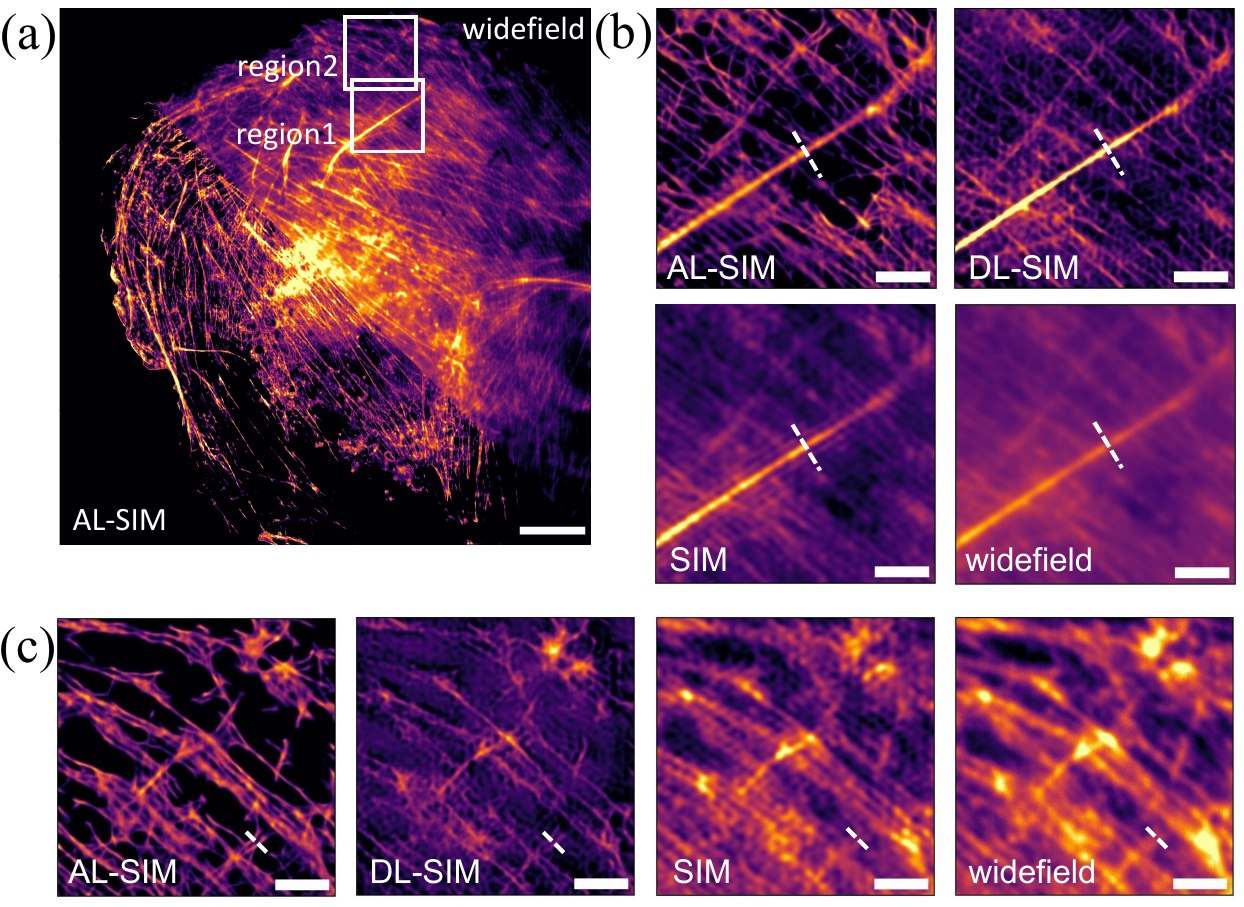}
        \includegraphics[width=0.9\linewidth]{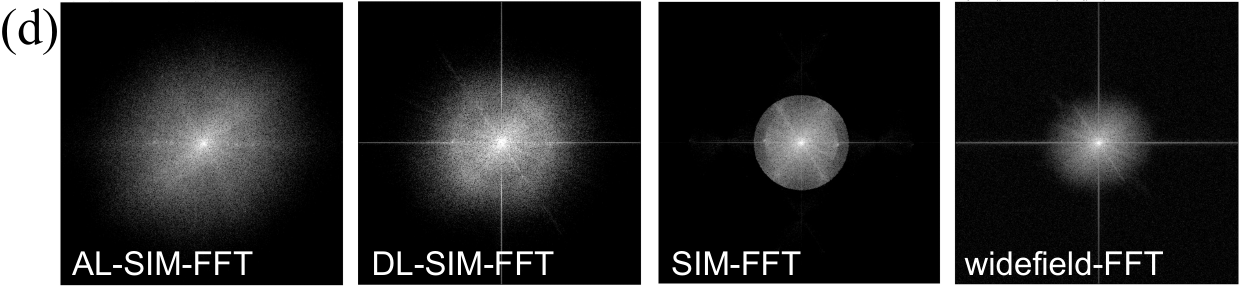}
        \includegraphics[width=0.9\linewidth]{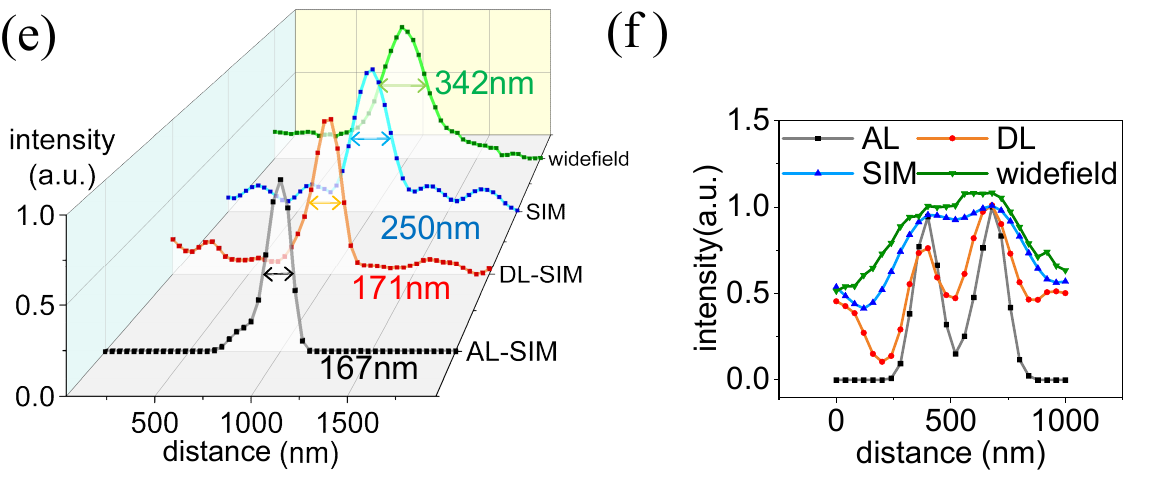}
        \caption{Comparative experimental results. (a) Comparison between AL-SIM and wide-field microscopy. (b) Imaging of region 1 using AL-SIM, DL-SIM, Conventional SIM, and wide-field microscopy. (c) Spatial resolution comparison along the white dashed line, highlighting that AL-SIM outperforms DL-SIM and conventional SIM in terms of artefacts. (d) Logarithmic Fourier spectra of imaging using AL-SIM, DL-SIM, Conventional SIM, and wide-field microscopy. (e) Imaging of region 1 using different techniques and the resolution improvement along the white dashed line. (f) Imaging of region 2 using different techniques and the resolution improvement along the white dashed line. (a: Scale: 10 $\mu$m; b, c: Scale: 2 $\mu$m)}
        \label{fig:result}
    \end{figure}

    In the real-world experiment, conventional SIM significantly improves the resolution of microfilament structures in BSC1 cells. However, as shown in Figs.~\ref{fig:result}(b-c), artefacts affect the accurate localisation of other structures located near areas of high fluorescence intensity. These artefacts arise from several factors, including inaccuracies in reconstruction parameters, movement of fluorescent molecules in living cells, refractive index mismatch and scattering in deeper biological tissues.

    The DL-SIM reconstruction method is trained on a large set of images produced by simulating the SIM imaging process, displaying robust generalization across various simulated sample types. However, when applied to real data, the training set does not account for the illumination field distortions encountered in practice, creating a mismatch between training and test conditions. Consequently, artefacts appear in several regions due to these unmodeled distortions in the illumination patterns, as shown in Figs.~\ref{fig:result}(b–c).

    The AL-SIM method estimates the light field and uses it to generate a simulated training set, aligning the training and test data and thus reducing artefacts (Fig.~\ref{fig:result}(c)). Comparing results from different methods reveals that in region~1, where fluorescent molecules are sparsely distributed, artefacts are significantly reduced. As shown in Fig.~\ref{fig:result}(e), these artefacts appear as side lobes; AL-SIM yields minimal side lobes. In region~2, densely packed fluorescent molecules result in superimposed artefacts from different molecules (Fig.~\ref{fig:result}(c)). Even under these high-density conditions, AL-SIM exhibits notable artifact suppression.

    Artefacts are often visible in the Fourier spectrum, manifesting as anomalous signals in the low-frequency region~\cite{pospivsil2017analysis, fan2019protocol}. Specifically, these artifact components appear as residuals in the Fourier domain~\cite{fan2019protocol}, as shown in Fig.~\ref{fig:result}(d). Side-by-side comparisons in Figs.~\ref{fig:result}(b–d) show that AL-SIM produces fewer artifact components than both DL-SIM and conventional SIM.

    We also perform resolution calibration, as shown in Fig.~\ref{fig:result}(e). Using the full width at half maximum (FWHM) method, we measure resolutions of 342 nm for wide-field, 250 nm for conventional SIM, 171 nm for DL-SIM, and 167 nm for AL-SIM. Additionally, using decorrelation analysis~\cite{descloux2019parameter}, we obtain values of 272.82 nm (wide-field), 308.31 nm (conventional SIM), 158.60 nm (DL-SIM), and 84.15 nm (AL-SIM). Notably, conventional SIM reconstruction fails to correctly calculate the decorrelation analysis resolution due to the influence of artefacts.

\section{Discussion}
     Compared to existing DL-based approaches, the AL-SIM method improves consistency between training and experimental data sets and reduces artefacts caused by out-of-distribution samples. It is applicable not only to cosine illumination light fields formed by double-beam interference, but also to more complex imaging techniques. However, it should be noted that the method may have difficulty detecting light fields in images with low contrast, low signal-to-noise ratio, and sparse distribution of fluorescent molecules (e.g., fluorescent beads).  We plan to extend this work by compensating for light fields in failed regions.

\section{Conclusion}
    In this work, we introduce AL-SIM, a novel approach to address the out-of-distribution problem in DL-SIM reconstructions. By accurately estimating the light field, AL-SIM effectively corrects errors due to data distribution variations, achieving twice the resolution of the diffraction limit and significantly reducing artefacts. Our method imprSoves the consistency between training and experimental data and enhances the applicability of SIM in complex biological environments.

\hfill

\noindent \textbf{Disclosures. } The authors declare no conflicts of interest.

\hfill

\bibliographystyle{unsrt}
\bibliography{ref}
\end{document}